\documentclass[twocolumn,english,aps,prb,superscriptaddress,showkeys,pdftex,reprint,longbibliography]{revtex4-1}

\usepackage[latin9]{inputenc}
\usepackage{verbatim}
\usepackage{amsmath}
\usepackage{amssymb}
\usepackage{graphicx}
\usepackage{babel}
\usepackage{mathtools}

\usepackage[labelfont=bf,font=small,justification=raggedright]{caption}
\DeclareCaptionLabelSeparator{Vert}{$~\vert~$}
\begin{document}

\title{Interacting polariton fluids in a monolayer of tungsten disulfide}

\author{Fábio Barachati}
\affiliation{Department of Engineering Physics, École Polytechnique de Montréal, Montréal H3C 3A7, QC, Canada}

\author{Antonio Fieramosca}
\affiliation{CNR - NANOTEC, Istituto di Nanotecnologia, Lecce 73100, Italy}

\author{Soroush Hafezian}
\affiliation{Department of Engineering Physics, École Polytechnique de Montréal, Montréal H3C 3A7, QC, Canada}

\author{Jie Gu}
\author{Biswanath Chakraborty}
\affiliation{Department of Physics, City University of New York, New York 10031, NY, USA}

\author{Dario Ballarini}
\email{dario.ballarini@gmail.com}
\affiliation{CNR - NANOTEC, Istituto di Nanotecnologia, Lecce 73100, Italy}

\author{Ludvik Martinu}
\affiliation{Department of Engineering Physics, École Polytechnique de Montréal, Montréal H3C 3A7, QC, Canada}

\author{Vinod Menon}
\affiliation{Department of Physics, City University of New York, New York 10031, NY, USA}

\author{Daniele Sanvitto}
\affiliation{CNR - NANOTEC, Istituto di Nanotecnologia, Lecce 73100, Italy}

\author{Stéphane Kéna-Cohen}
\email{s.kena-cohen@polymtl.ca}
\affiliation{Department of Engineering Physics, École Polytechnique de Montréal, Montréal H3C 3A7, QC, Canada}

\date{\today}

\maketitle

\textbf{Atomically thin transition metal dichalcogenides (TMDs) possess a number of properties that make them attractive for realizing room-temperature polariton devices\cite{mak2016photonics}. An ideal platform for manipulating polariton fluids within monolayer TMDs is that of Bloch surface waves, which confine the electric field to a small volume near the surface of a dielectric mirror\cite{doi:10.1063/1.3571285,Lerario14,doi:10.1063/1.4863853,lerario2017high}. Here we demonstrate that monolayer tungsten disulfide ($\text{WS}_2$) can sustain Bloch surface wave polaritons (BSWPs) with a Rabi splitting of 43 meV and propagation constants reaching 33 $\boldsymbol{\mu}$m. In addition, we evidence strong polariton-polariton nonlinearities within BSWPs, which manifest themselves as a reversible blueshift of the lower polariton resonance by up to 12.9$\pm$0.5 meV. Such nonlinearities are at the heart of polariton devices\cite{Ciuti2003,amo2009collective,amo2010exciton,PhysRevB.87.195305,sturm2014all,daskalakis2014nonlinear,sanvitto2016road} and have not yet been demonstrated in TMD polaritons. As a proof of concept, we use the nonlinearity to implement a nonlinear polariton source. Our results demonstrate that BSWPs using TMDs can support long-range propagation combined with strong nonlinearities, enabling potential applications in integrated optical processing and polaritonic circuits.}

Strong light-matter coupling with TMD excitons has previously been demonstrated in a variety of systems, including planar microcavities\cite{dufferwiel2015exciton,liu2015strong,flatten2016room,6b014752016,sun2017optical} and plasmonic cavities\cite{6b014752016,5b045882016,lundt2016room}. At room-temperature, however, these structures have been limited to extremely short polariton lifetimes due to the high losses of the underlying cavity and the broad exciton linewidth. Here, we overcome these limitations by strongly coupling the A exciton of monolayer $\text{WS}_2$ to a low-loss propagating Bloch surface wave at the air-dielectric interface of a Bragg mirror. Tungsten disulfide was chosen as the active material over other TMDs due to its strong and sharp excitonic absorption, which better matches the narrow linewidths of Bloch surface modes. The low losses of the dielectrics and the thin monolayer enable polaritons to propagate over the entire extent of the monolayer. To highlight the role of polariton-polariton nonlinearities, we demonstrate a nonlinear polariton source in a configuration analogous to that previously used to show bistability under continuous wave excitation\cite{PhysRevA.69.023809}. All measurements reported here were performed under ambient conditions.

\begin{figure}[ht]
\includegraphics[width=0.45\textwidth]{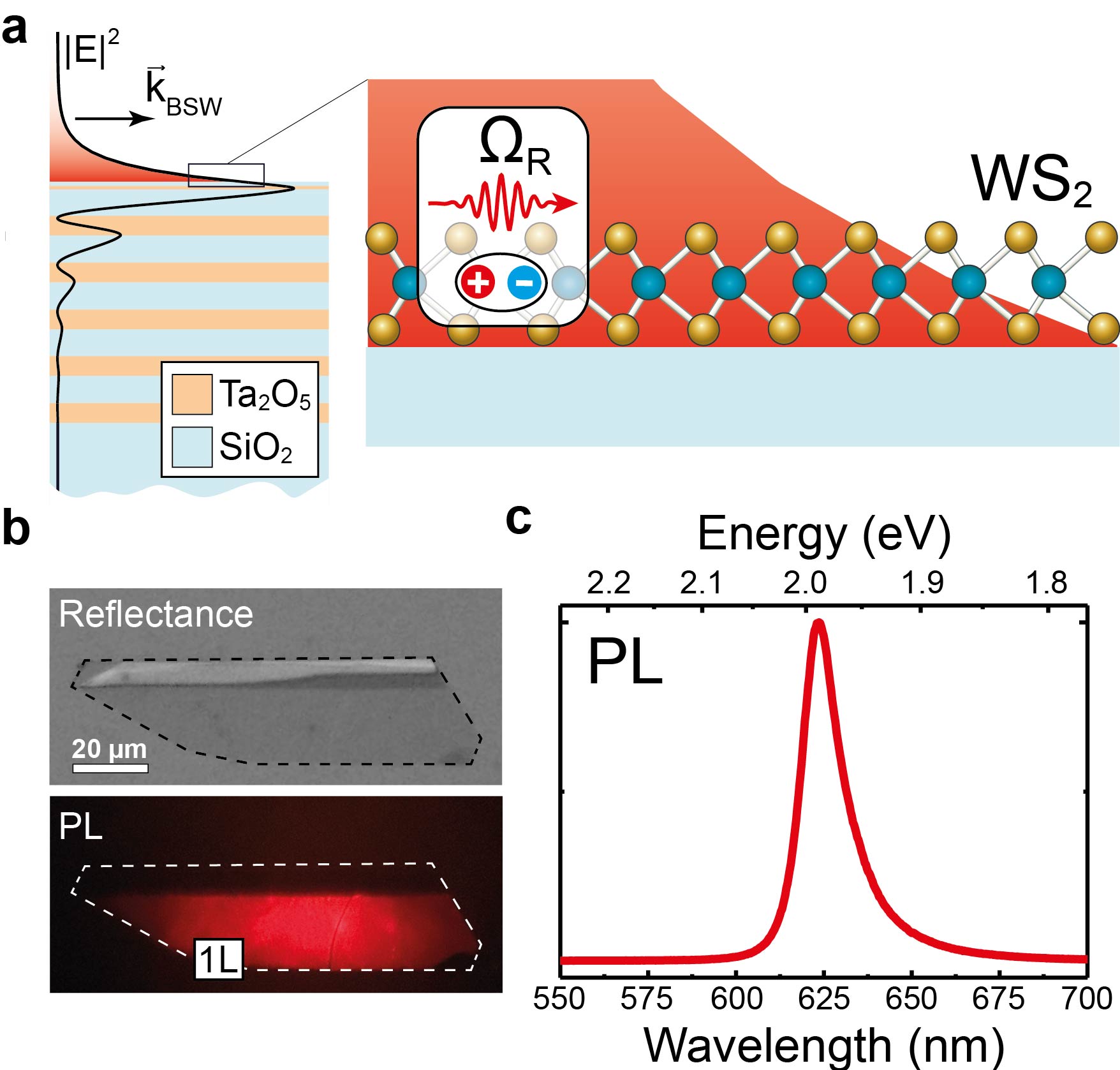}
\caption{\textbf{Sample structure and monolayer characterization.} {\normalfont(}\textbf{a}{\normalfont)} {\normalfont Schematic of the dielectric stack supporting Bloch surface wave polaritons in monolayer $\text{WS}_2$. The solid black lines trace the electric field profile of the bare mode at the wavelength of the A exciton band. The mode is TE polarized and propagates along the surface with wavevector $\vec{k}_{BSW}$. The inset illustrates the coupling of the enhanced electric field at the surface of the stack to the in-plane excitons in the monolayer. }{\normalfont(}\textbf{b}{\normalfont)} {\normalfont Micrographs of the monolayer in reflectance and PL. The dashed lines indicate the flake boundaries. Only monolayer regions show bright PL under illumination by a large Gaussian spot. }{\normalfont(}\textbf{c}{\normalfont)} {\normalfont The typical monolayer PL spectrum under 514 nm excitation contained a single strong peak centered at 1.988 eV/623.6 nm.}}
\label{FigSample} 
\end{figure} 

\begin{figure*}[ht]
\includegraphics[width=1\textwidth]{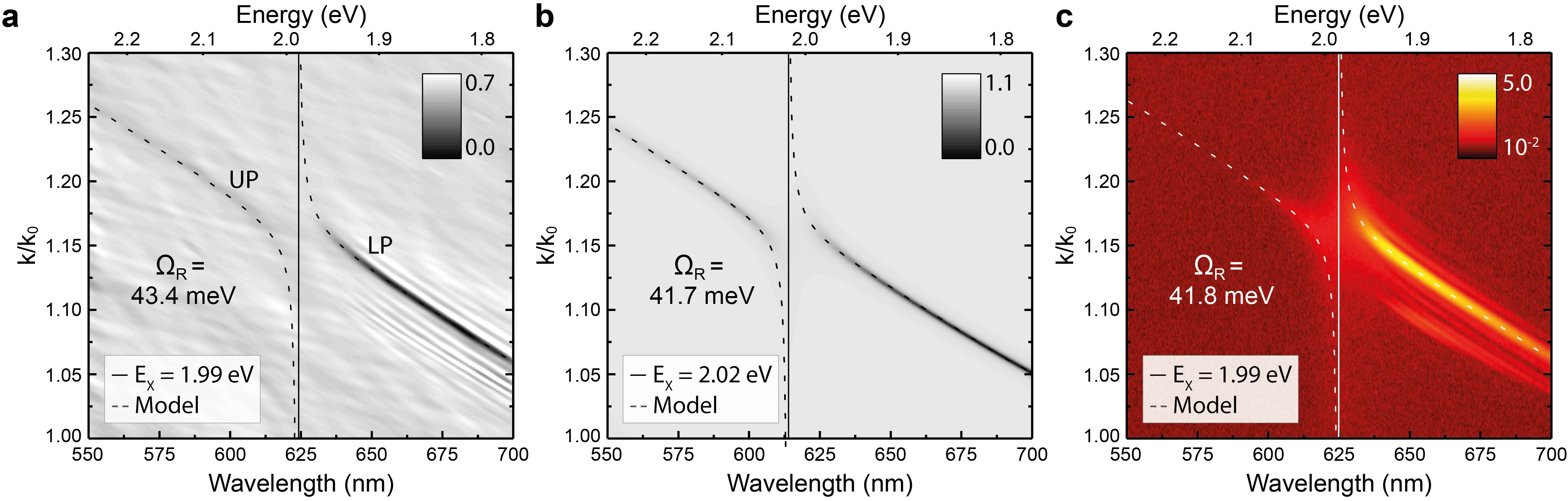}
\caption{\textbf{Bloch surface wave polaritons.} {\normalfont(}\textbf{a}{\normalfont)} {\normalfont Experimental dispersion of the Bloch surface wave polaritons in a monolayer $\text{WS}_2$ in reflectance and }{\normalfont(}\textbf{b}{\normalfont)} {\normalfont the corresponding transfer matrix calculation. The experimental and theoretical dispersions for the bare Bloch mode are shown in Fig. S1.} {\normalfont(}\textbf{c}{\normalfont)} {\normalfont Experimental dispersion in PL under 514 nm excitation on a logarithmic color scale. Coupled harmonic oscillator fits and fitting parameters are shown in all three panels. The UP and LP modes are traced in dashed lines and the fitted exciton energies are indicated by solid lines.}}
\label{FigWS2} 
\end{figure*}

A schematic of our sample is shown in Fig.~\ref{FigSample}a. A glass coverslip was coated with a dielectric mirror, designed to support a Bloch surface mode near the A exciton band of $\text{WS}_2$ (2.014 eV/615.6 nm)\cite{PhysRevB.90.205422}. A large monolayer of $\text{WS}_2$ was then transferred onto the top dielectric surface. The solid line in Fig.~\ref{FigSample}a shows the calculated electric field profile corresponding to the bare Bloch mode at the A exciton wavelength. The field peaks inside the last dielectric pair and decays exponentially away from the surface. The mode is TE polarized and propagates along the surface with wavevector $\vec{k}_{BSW}$. Fig.~\ref{FigSample}b shows a micrograph of the large exfoliated $\text{WS}_2$ flake in reflectance (top) and in photoluminescence (PL, bottom) under 514 nm excitation by a large Gaussian spot. Only the monolayer regions exhibit bright PL due to their direct bandgap\cite{doi:10.1021/nl3026357}. A typical monolayer PL spectrum is shown in Fig.~\ref{FigSample}c and contains a single strong peak centered at 1.988 eV/623.6 nm with a FWHM of 42 meV. These values vary slightly along the sample, presumably due to fluctuations in strain, substrate adhesion, defects and surface charge density\cite{zhu2015exciton,ADOM:ADOM201700767}.

To demonstrate strong coupling between the monolayer A exciton and the Bloch surface mode, white-light reflectivity was measured with an immersion objective in a back focal plane (BFP) imaging configuration (see Methods). In Fig.~\ref{FigWS2}a we show the experimental dispersion of the upper (UP) and lower (LP) polariton modes measured in the center of the monolayer. The position and visibility of the modes are in good agreement with transfer matrix calculations shown in Fig.~\ref{FigWS2}b, where the thickness and refractive index of monolayer $\text{WS}_2$ were obtained from the literature\cite{PhysRevB.90.205422}. Both polariton branches were also visible in PL, as shown in Fig.~\ref{FigWS2}c on a logarithmic color scale. Their characteristic anti-crossing is visible in both reflectance and PL around the same wavelength of 623 nm, coinciding with the peak PL wavelength shown in Fig~\ref{FigSample}c. Interestingly, a progression of modes surrounding the LP branch is visible in Fig.~\ref{FigWS2}a,c. Bloch surface waves are extremely sensitive to changes in the thickness and refractive index of the topmost layer. In the case of our monolayer, these can be caused by surface inhomogeneities in the large area probed by the propagating mode. A similar but weaker effect can also be seen for the bare mode, shown in Fig. S1a. For each panel in Fig.~\ref{FigWS2}, a simple 2$\times$2 coupled harmonic oscillator (CHO) model was used to fit the data (see Methods). The dispersion fits and exciton energies are traced in dashed and solid lines, respectively. The extracted Rabi splittings of 43.4$\pm$0.8 meV and 41.8$\pm$0.6 meV are in close agreement with the transfer matrix value of 41.7$\pm$ 0.3 meV. 

\begin{figure*}[t]
\includegraphics[width=1\textwidth]{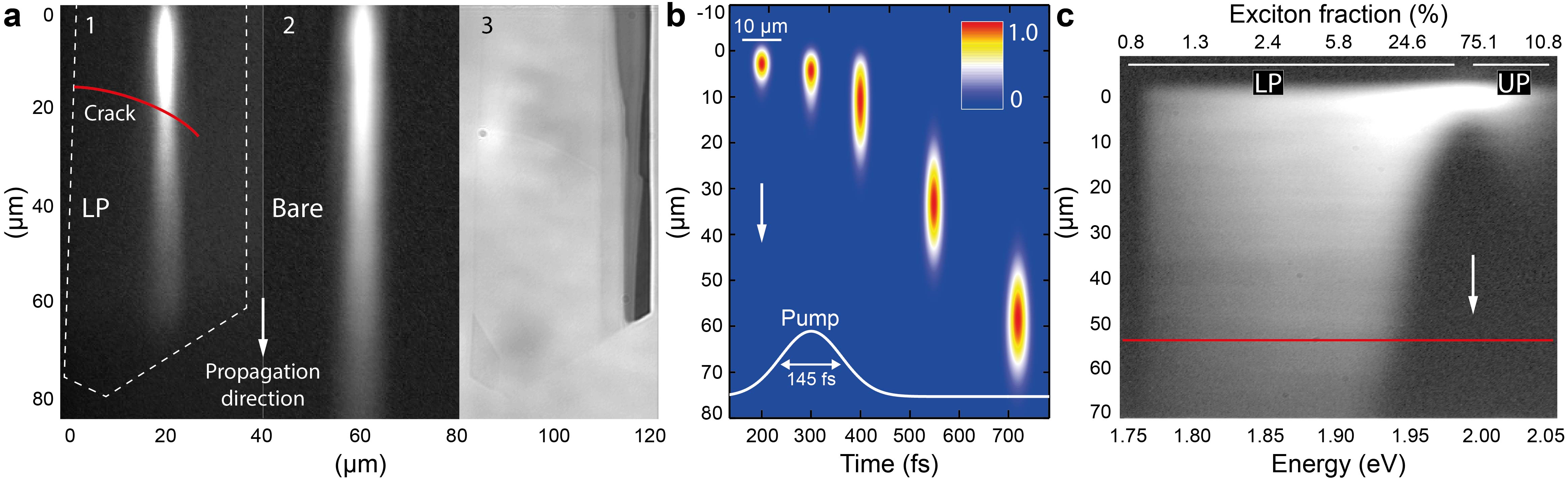}
\caption{\textbf{Polariton propagation.} {\normalfont(}\textbf{a}{\normalfont)} {\normalfont Resonant propagation at a wavelength of 645 nm for the LP (exciton fraction of 10\%) and bare modes, shown in panels 1 and 2, respectively. Corresponding propagation constants are 20.6$\pm$0.1$~\mu$m and 21.1$\pm$0.1$~\mu$m. The flake boundary is shown by a white dashed line and a small crack is indicated by a solid red line. Panel 3 shows a micrograph of the monolayer in reflectance where the crack and boundaries can be seen. }{\normalfont(}\textbf{b}{\normalfont)} {\normalfont Calculated time snapshots of the spatial density of BSWPs following the arrival of a laser pulse, traced in a solid white line. The chirped pump pulse has an estimated FWHM of 145 fs. The polariton wavepacket is initially concentrated close to the excitation spot. As the pump vanishes, it can be seen propagating downwards at a group velocity of 1.49$\cdot10^8~\text{m}~\text{s}^{-1}$. The propagation traces were displaced laterally so that the center of the excitation spot coincides with the time at which the snapshot was taken. The top scale bar is related to the spatial dimensions of the propagation traces in the horizontal direction.}{\normalfont(}\textbf{c}{\normalfont)} {\normalfont Non resonant (PL) propagation under 514 nm excitation on a logarithmic intensity scale. The solid red line indicates the position of a small crack on the flake. Propagation constants for the LP branch ranged from 14.7$\pm$0.1 $\mu$m to 33.4$\pm$0.4 $\mu$m for exciton fractions between 15.7\% to 1.7\%. Polaritons from the upper branch are also visible but propagate significantly less. In both measurements the propagation direction is downwards, as indicated by the white arrows.}}
\label{FigProp} 
\end{figure*}

Next, we investigated how far BSWPs are able to propagate within the $\text{WS}_2$ monolayer. The pump wavevector and wavelength were selected to be in resonance with the LP mode. The first panel in Fig.~\ref{FigProp}a shows the real space propagation trace for an exciton fraction of 10\% and wavelength of 645 nm. Propagation can be observed for over 60 $\mu$m and ends upon reaching the flake boundary, indicated by a dashed white line. As a comparison, the propagation of the uncoupled (bare) Bloch surface wave is shown  in panel 2. The propagation constants extracted from single exponential decay fits were 20.6$\pm$0.1$~\mu$m and 21.1$\pm$0.1$~\mu$m for the LP and bare modes, respectively. These values are considerably larger than the ones found in high-quality planar microcavities embedding TMD monolayers, which are typically of the order of 1 $\mu$m\cite{liu2015strong}. Here, the propagation length can be further increased by limiting the angular content of the excitation beam. By reducing divergence in this way, the propagation length of the bare mode could be increased to 42.2$\pm$0.2$~\mu$m (Fig. S3). We observe that the presence of a small crack in the flake, indicated by a solid red line in panel 1, has very little impact on the propagation, possibly due to the high photonic content of the mode. The third panel in Fig. \ref{FigProp}a shows an enlarged micrograph of the monolayer in reflectance where the small crack and monolayer boundaries can be seen. Given that these measurements were performed using pulsed excitation, we illustrate in Figure~\ref{FigProp}b the underlying temporal dynamics by calculating the spatial density of BSWPs under our experimental conditions (see Methods and the Supplementary Information). The temporal profile of the pump is traced in a solid white line. During the pump pulse, the density of polaritons is highest close to the excitation spot, centered at $\mathbf{r}=0$. As the pump vanishes, the polariton wavepacket can be clearly seen as it propagates downwards with a group velocity of 1.49$\cdot10^8~\text{m}~\text{s}^{-1}$. 

Propagation was also investigated using non-resonant above-gap excitation. In this case, the pump first creates excitons, which subsequently relax into propagating polariton states. Fig.~\ref{FigProp}c shows the PL spectrum under 514 nm excitation as it propagates within the flake. The propagation constants for polaritons with exciton fractions ranging from 1.7\% to 15.7\% were found to be between 33.4$\pm$0.4 $\mu$m and 14.7$\pm$0.1 $\mu$m, consistent with the resonant case. Polaritons from the upper branch are also visible in Fig.~\ref{FigProp}c due to the logarithmic intensity scale, but propagate significantly less. Again, the presence of the small crack is indicated by the solid red line and only a negligible effect was seen on the intensity profiles. The single-exponential fits for both resonant and non-resonant propagation traces are shown in Fig. S2.

The experimental LP propagation constant for an exciton fraction of 15.7\% agrees well with the value of 15.3 $\mu$m, estimated from the group velocity and a 6.4 meV linewidth (103 fs lifetime), extracted from the reflectivity spectra. The maximum propagation constant that can be estimated from the dispersion relation is bound by our detection limit of 5 meV to 19.6 $\mu$m. From the experimental propagation constants we obtain a linewidth of 2.4 meV for the bare mode, corresponding to a quality factor $Q = 800$ at a wavelength of 645 nm.

\begin{figure*}[ht]
\includegraphics[width=1\textwidth]{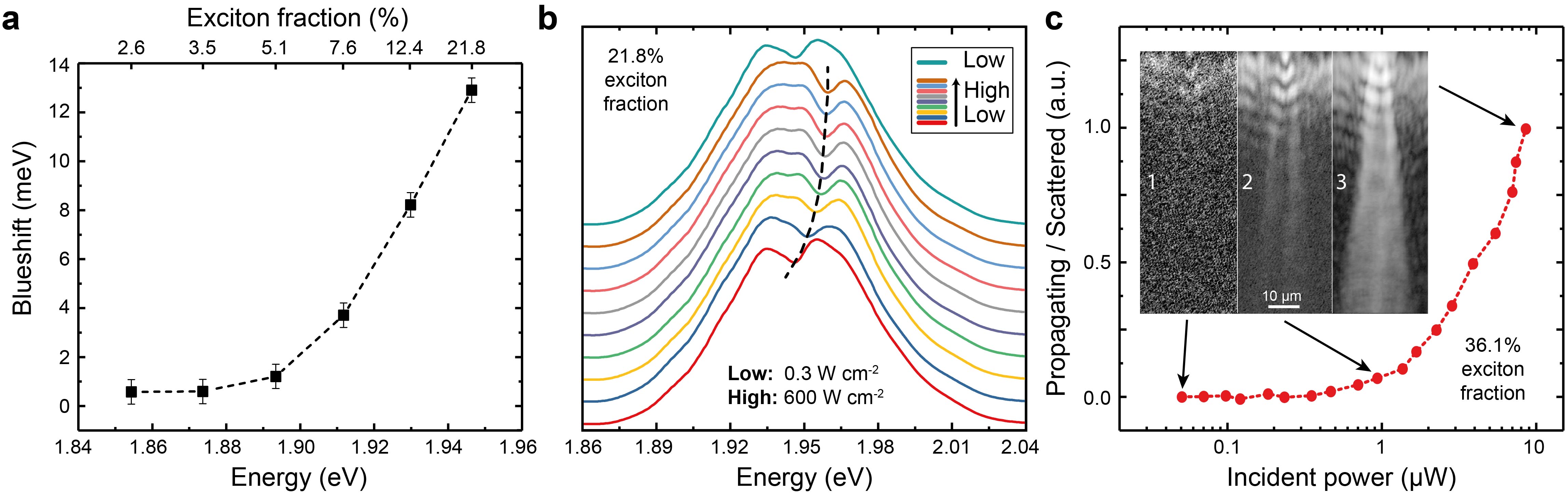}
\caption{\textbf{Nonlinear behaviour.} {\normalfont(}\textbf{a}{\normalfont)} {\normalfont High power (600 W cm$^{-2}$) time-averaged resonant blueshift of the LP mode as a function of the initial BSWP energy and corresponding exciton fraction, extracted from the CHO model shown in Fig.~\ref{FigWS2}a. }{\normalfont(}\textbf{b}{\normalfont)} {\normalfont Power sweep in steps of $75~\text{W cm}^{-2}$ for the highest blueshift/exciton fraction, showing that the shift is reversible and larger than the LP linewidth of 7$\sim$8 meV. }{\normalfont(}\textbf{c}{\normalfont)} {\normalfont A nonlinear polariton source, where the incident power can block or launch a beam of propagating BSWPs. The red curve shows the normalized ratio of the propagating and scattered beams, which depend on their integrated areas, illustrating the nonlinear dependence on the incident pump power.}}
\label{FigBlue} 
\end{figure*}

Next, we study the effect of polariton-polariton interactions. At high densities, polaritons interact through their matter component due to phase space filling (PSF) and inter-particle Coulombic interactions, which under most conditions are repulsive and lead to a blueshift of the polariton modes. Fig.~\ref{FigBlue}a shows the resonant blueshift of the LP mode at the highest incident power density ($600~\text{W cm}^{-2}$) as a function of the resonance position in the linear regime and the exciton fraction extracted from the CHO model shown in Fig.~\ref{FigWS2}a. In Fig.~\ref{FigBlue}b we show the complete power sweep where the highest time-averaged blueshift of 12.9$\pm$0.5 meV was observed. The top two curves for low and high powers show that the blueshift is reversible and is larger than the LP linewidth of 7$\sim$8 meV. The LP dispersions at low and high pump powers are shown in Fig. S4. The blueshift saturates with power (Fig. S5), as evidenced by flattening out of the dashed line in Fig.~\ref{FigBlue}b at higher powers. The saturation mechanism is likely due to a combination exciton-exciton annihilation and a possible dynamic transition to weak coupling at the highest powers\cite{C5NR00383K}.

Estimating the polariton-polariton interaction constants from the experimental LP blueshift requires an accurate knowledge of the polariton density. In addition, when several modes in momentum space are excited, their individual contributions to the blueshift must be accounted for\cite{Ciuti2003}. Using input-output theory and parameters corresponding to our experimental conditions, we have performed time-dependent calculations of the polariton density per mode in momentum space (see Methods and Fig. S6). Over the range of momenta probed, our experiments cannot determine the relative contribution of each interaction mechanism (Fig. S7). However, we can separately estimate the required exciton-exciton $(V_{XX})$ and PSF $(V_{SAT})$ interaction constants required to explain the observed blueshift. We find $V_{XX}=0.5\pm0.2~\mu\text{eV}~\mu\text{m}^2$ and $V_{SAT}=0.09\pm0.04~\mu\text{eV}~\mu\text{m}^2$. Remarkably, these values are considerably higher than those reported for other room-temperature systems\cite{PhysRevLett.115.035301,lerario2017room}.

Our interaction constants can also be compared to theoretical values. The  exciton-exciton interaction constant in TMDs can be estimated using $V_{XX}\simeq 2.07E_{1s}\lambda_X^2=1.9~\mu\text{eV}~\mu\text{m}^2$, where $E_{1s}=0.32~\text{eV}$ is the 1s exciton binding energy and $\lambda_X=1.7~\text{nm}$ is the 1s exciton radius\cite{PhysRevB.96.115409}. The  numerical prefactor in the expression for PSF depends on the form of the exciton wavefunctions. For 1s excitons, it is often estimated\cite{Ciuti2003} as $V_{SAT}\simeq 7.18\hbar\Omega_R\lambda_X^2=0.87~\mu\text{eV}~\mu\text{m}^2$. Both theoretical values slightly exceed those obtained from the blueshift. This is, however, consistent with the fact that peak densities were used in our estimates, but experimental time-averaging will reduce the apparent blueshift. These estimates also suggest that both Coulombing interactions and PSF play a comparable role within the range of momenta probed in our system.

Finally, we demonstrate how the optical control of the LP mode can be used as a nonlinear source of polaritons. The pump bandwidth was limited to 3 nm centered at 633 nm and at this wavelength the LP resonance is too broad and shallow to be resolved in Fig.~\ref{FigWS2}a. At low incident powers, very little light is coupled into the propagating polariton mode, as shown on inset 1 in Fig.~\ref{FigBlue}c. As the pump power is increased, the LP mode blueshifts and the propagating part of the BSWP dispersion moves closer to resonance with the pump (insets 2 and 3). At high power, the pump laser is fully resonant with the polariton mode and launches a propagating surface wave, which through coupling to the underlying bare mode is able to propagate far beyond the flake dimensions. The normalized ratio of the propagating and scattered powers, which depends on the selected integration areas, is shown in the red curve, illustrating the nonlinear dependence on the incident pump power. This behaviour is analogous to the onset of bistability, which can be observed for continuous wave pumping in planar microcavities\cite{PhysRevA.69.023809}.

Our demonstration of strong light-matter coupling between excitons in two-dimensional materials and propagating Bloch surface waves introduces an exciting new platform for the study of interacting polariton fluids at room temperature. Polariton propagation losses stem mainly from leakage into the immersion optics, which could be reduced by an increase in the number of dielectric pairs or even completely avoided by the use of surface gratings for external in- and out-coupling\cite{angelini2014focusing}. The simple dielectric structure can be tailored for other two-dimensional materials or for multilayer heterostructures. In particular, TMDs encapsulated by hexagonal boron nitride show improved surface flatness, screening from charged impurities and ambient stability. The reduced disorder can lead to a suppression of loss mechanisms such as exciton-exciton annihilation and allow higher polariton densities to be achieved in the nonlinear regime\cite{PhysRevB.95.241403}. Interacting BSWPs based on TMDs can enable the fabrication of room-temperature nonlinear polariton devices with long propagation distances. Together with properties such as high photostability, valley polarization and tunability by electric fields, TMDs could also bring new functionalities to polaritonic circuits.

\section*{Methods}

The Bragg mirror consisted of five pairs of tantalum pentoxide ($\text{Ta}_2\text{O}_5$)/silicon dioxide ($\text{SiO}_2$) layers (98.5 nm/134.6 nm thick), deposited by radiofrequency magnetron sputtering at a pressure of $10^{-7}$ mbar. An additional thinner pair (17.1 nm/22.3 nm) was used to shift the position of the Bloch mode at the tungsten disulfide ($\text{WS}_2$) A exciton wavelength towards the center of the photonic bandgap. A large monolayer of $\text{WS}_2$ was first tape-exfoliated onto a polydimethylsiloxane stamp and subsequently transferred onto the top dielectric surface. 

The experimental setup for reflectivity measurements consisted of an inverted microscope (Olympus IX-81) equipped with a 1.42 numerical aperture (NA) oil immersion objective. At the microscope's side port, three two-inch achromatic doublet lenses (Thorlabs, focal lengths 20/7.5/30 cm) were used to project a magnified image of the back focal plane (BFP) onto the slit of an imaging spectrometer with a cooled CCD camera (Princeton Instruments, IsoPlane 160 spectrometer, PIXIS 400B eXcelon camera, 300 g/mm grating with 500 nm blaze, 50 $\mu$m slit). The white light source (Energetiq, EQ-99X) was spatially filtered by a single-mode fiber (Thorlabs P1-630A-FC) and a large area Gaussian collimator (SLT, LB20). The same setup was used for photoluminescence (PL) and non-resonant propagation measurements by using a CW 514 nm diode laser (Thorlabs L520P50/LTC56B) as the light source, coupled to a Thorlabs P1-405B-FC fiber. The filters used were Omega Filters RPE520SP (excitation), Semrock FF538-FDi01 (dichroic) and Thorlabs FELH0550 (detection). 

For resonant propagation and nonlinear measurements, a tunable femtosecond laser (estimated pulse width of 145 fs, repetition rate 10 kHz) was focused onto the BFP of a 1.49 NA oil immersion objective with a long focal length lens (75 cm). A home built microscope was used with the same lenses as the setup above, except for a 20 cm tube lens. The energy and focusing position of the laser were adjusted to be resonant with the mode of interest and the corresponding real space spot size dimensions were typically 3$~\mu$m $\times$ 5$~\mu$m (full-width at half-maximum). The same spectrometer was used. Both the bare Bloch surface wave and the Bloch surface wave polariton modes were visible in reflectance close to the boundaries of the monolayer due to the reduced coupling strength. During propagation, polaritons approaching the boundary leaked light into the underlying bare mode and appeared to propagate beyond the boundary of the flake with a small change in propagation direction. For propagation measurements, adjustable slits were placed in the BFP to select only the lower polariton (LP) mode and in the image plane to block the excitation spot (in the resonant case only).

Reflectivity and PL spectra were fitted with the simple 2$\times$2 coupled harmonic oscillator Hamiltonian
\begin{equation}
\hat{H}_{\mathbf{k}}=\begin{pmatrix}E_{BSW}(\mathbf{k}) & \Omega_R/2\\
\Omega_R/2 & E_{EX}
\end{pmatrix},
\label{EqHBSW}
\end{equation}
where $\Omega_R$ is the Rabi splitting, $E_{EX}$ is the exciton energy and $E_{BSW}(\mathbf{k})$ is the energy dispersion of the bare Bloch surface mode, which can be approximated by
\begin{equation}
E_{BSW}(\mathbf{k})=\hbar v_g|\mathbf{k}|+E_0
\label{EqBSW}
\end{equation}
in the center of the photonic bandgap. Here, $v_g$ is the group velocity and $E_0$ is a fitting parameter.

For numerical simulations, we calculate the LP field in momentum space $\psi_{LP}(\mathbf{k},t)$. Its time-evolution is governed by 
\begin{multline}
i\hbar \frac{\partial\psi_{LP}(\mathbf{k},t)}{\partial t} = \left[E_{BSWP}(\mathbf{k}) - \hbar\omega_P - \frac{i\hbar\gamma_{LP}}{2}\right]\psi_{LP}(\mathbf{k},t)\\ + \hbar P(\mathbf{k},t),
\end{multline}
where $E_{BSWP}(\mathbf{k})$ is a linear approximation of the LP dispersion in the vicinity of the pump, $\hbar\omega_P$ is the pump energy, $\gamma_{LP}$ is the LP dissipation rate and $P(\mathbf{k},t)$ is the driving term. To reproduce our experimental conditions, the pump field is taken to be a Gaussian in momentum space modulated by a positively chirped temporal Gaussian envelope. 

The interaction constants were calculated at an incident pump power below the saturation of the blueshift (150 W cm$^{-2}$) and at the instant of highest total polariton density in momentum space. The $\mathbf{k}$ dependent blueshift $\Delta E_{LP}(\mathbf{k})$ is given by
\begin{equation}
\Delta E_{LP}(\mathbf{k}) = \sum_{\mathbf{k'}}E_{\mathbf{k},\mathbf{k'}}^{\text{shift}}|\psi_{LP}(\mathbf{k'})|^2,
\end{equation}
where $E_{\mathbf{k},\mathbf{k'}}^{\text{shift}}$ is the polariton-polariton interaction constant\cite{Ciuti2003}. Substituting the expressions for each interaction mechanism individually (see Supplementary Information), we can obtain the exciton-exciton interaction constant as
\begin{equation}
V_{XX}(\mathbf{k}) = \frac{\Delta E_{LP}(\mathbf{k})}{2|X_{\mathbf{k}}|^2\sum_{\mathbf{k'}}|X_{\mathbf{k'}}|^2|\psi_{LP}(\mathbf{k'})|^2}
\end{equation}
and the saturation interaction constant as
\begin{equation}
V_{SAT}(\mathbf{k}) = \frac{\Delta E_{LP}(\mathbf{k})}{\splitfrac{|C_{\mathbf{k}}||X_{\mathbf{k}}|\sum_{\mathbf{k'}}{|X_{\mathbf{k'}}|^2|\psi_{LP}(\mathbf{k'})|^2}}{+3|X_{\mathbf{k}}|^2\sum_{\mathbf{k'}}{|C_{\mathbf{k'}}||X_{\mathbf{k'}}||\psi_{LP}(\mathbf{k'})|^2}}},
\end{equation}

where $|C_{\mathbf{k}}|$ and $|X_{\mathbf{k}}|$ are the Hopfield coefficients for the LP photon and exciton fractions, respectively. All the expressions and parameter values used in the calculations are listed in the Supplementary Information.

\bibliography{References}

\begin{acknowledgments}
SKC acknowledges support from the NSERC Discovery Grant Program and the Canada Research Chair in Hybrid and Molecular Photonics. FB acknowledges support from the FQRNT PBEEE scholarship program. LM acknowledges financial support from the NSERC Discovery Grant. Work at the City University of New York was supported by the National Science Foundation (NSF) under the EFRI 2-DARE program (EFMA-1542863) and the NSF-ECCS-1509551 grant. AF, DB and DS acknowledge the ERC "ElecOpteR" grant number 780757. SKC and DS acknowledge support from the mixed Québec-Italy sub-commission for bilateral collaboration.
\end{acknowledgments}

\section*{Author contributions}
SKC conceived the project and FB designed the sample. The sample was fabricated by FB, SH, JG, BC under the supervision of LM, SKC and VM. Optical experiments were performed by FB, AF and DB. FB analyzed the data and wrote the manuscript. Numerical calculations were performed by FB and SKC. All authors contributed to revising the manuscript and analyzing the results. DB, DS and SKC coordinated the project. 

\section*{Competing financial interests}
The authors declare no competing financial interests.

\end{document}